\documentclass[aps,prl,showpacs,superscriptaddress,twocolumn]{revtex4}

\usepackage{pifont}
\usepackage{graphicx}
\usepackage{amssymb}
\usepackage{epsfig}
\usepackage{amsmath}
\usepackage{dcolumn}
\usepackage{indentfirst}
\usepackage{natbib}
\usepackage{CJK}  
\begin{document}
\begin{CJK*}{UTF8}{gbsn}
\bibliographystyle{plainnat}
\author{Zhu Chen (陈竹)}
\affiliation{Institute of Physics, Chinese Academy of Sciences, 100190, Beijing, China}
\author{Biao Wu (吴飙)}
\affiliation{International Center for Quantum Materials, Peking University, 100871, Beijing, China}
\title{Bose-Einstein Condensate in a Honeycomb Optical Lattice:
Fingerprint of Superfluidity at the Dirac Point}
\begin{abstract}
Mean-field Bloch bands of a Bose-Einstein condensate in a honeycomb optical
lattice are computed. We find that the topological structure of the Bloch bands
at the Dirac point is changed completely by atomic interaction of
arbitrary small strength: the Dirac point is extended into a closed curve and an
intersecting tube structure arises around the original Dirac point.
These tubed Bloch bands are caused by the superfluidity of the system.
Furthermore, they imply the inadequacy of the tight-binding model to describe
an interacting Boson system around the Dirac point and the breakdown of
adiabaticity by interaction of arbitrary small strength.
\end{abstract}
\date{\today}

\maketitle

\end{CJK*}

Inspired by the exciting physics in graphene\cite{novoselov, ybzhang, wu, wyao}, there
have been increasing efforts to study ultracold atoms in a honeycomb optical lattice
\cite{Zhu,Ablowitz,Treidel,peleg,Panahi,haddad,carr}.
The primary reason is that these ultracold atom systems offer more
controlling flexibilities over graphene\cite{Morsch,Yukalov}.
For example, with this hexagonal ultracold atom system,  one can readily
change the lattice strength, tune the atomic scattering strength with Feshbach resonance,
and load either bosons or fermions or even a mixture of bosons and fermions in the lattice.
There have already been efforts to study conical diffraction\cite{Ablowitz,Treidel,peleg}
and observe quantum phases with ultracold bosons in a honeycomb lattice\cite{Panahi}.
This controlling flexibility will not only offer deeper insight into the graphene properties
but also open up windows for physics beyond graphene.

In this Letter we provide an insight into the interplay between superfluidity and
Dirac dynamics by studying a Bose-Einstein condensate (BEC) in a
honeycomb optical lattice.  To showcase the interplay, we compute the lowest
Bloch bands for this BEC system. We find that the topology of the Bloch bands
around the Dirac point is completely altered by arbitrary small atomic interaction:
an intersecting tube  structure appears  and the Dirac point is turned into a closed curve.
We show that the topological change can be viewed as a permanent fingerprint
left in the Bloch bands by superfluidity.  As the interaction does not change the
Dirac point structure in the tight-binding model,
this topological change suggests that the tight-binding model is insufficient
to describe the bosonic dynamics in a honeycomb lattice no matter how deep the lattice is.
At the same time, these tubed bands imply the breakdown of adiabaticity by arbitrary small
atomic interaction. A feasible experimental scheme is suggested to observe
this phenomenon.
\begin{figure}
\includegraphics[angle=0,width=7.5cm]{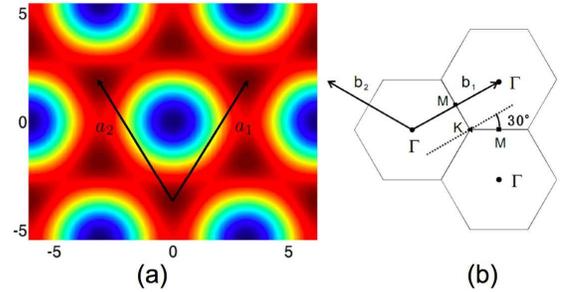}
\caption{(color online) (a) Contour map of the hexagonal potential in Eq.(\ref{eqv}).
The potential well is represented in red and the barrier in blue.
The unit vectors are marked as $\mathbf{a}_{1}$ and $\mathbf{a}_{2}$.
(b) Unit cell in the reciprocal space with unit vectors $\mathbf{b}_{1}$, $\mathbf{b}_{2}$
and the high symmetry points $\mathbf{\Gamma}$, $\mathbf{M}$ and $\mathbf{K}$.
}
 \label{hconfig}
\end{figure}

The honeycomb optical lattice can be experimentally realized by three interfering
traveling laser beams\cite{grynberg,wunsch}, and is described mathematically by
\begin{equation}
\label{eqv}
V(\mathbf{r})=V_{0}\Big[\cos(\mathbf{b}_{1}\cdot \mathbf{r})+
\cos(\mathbf{b}_{2}\cdot \mathbf{r})+\cos((\mathbf{b}_{1}+\mathbf{b}_{2})\cdot \mathbf{r})\Big],
\end{equation}
where the reciprocal unit vectors $\mathbf{b}_{1}=2\pi (\sqrt{3}, 1)/(3a)$ and
$\mathbf{b}_{2}=2\pi (-\sqrt{3}, 1)/(3a)$ with $a=2\lambda_L/3\sqrt{3}$.
$\lambda_L$ is the wavelength of the laser beams. We are interested in the superfluid
regime, where the BEC system can be well described by the Gross-Pitaevskii (GP) equation
\begin{equation}
\label{gpe}   i\hbar\frac{\partial \psi}{\partial
t}=-\frac{\hbar^{2}}{2m}\nabla^{2}\psi+V(\mathbf{r})
\psi+\frac{4\pi\hbar^{2}a_{s}}{m}\left|\psi\right|^{2}\psi.
\end{equation}
with $m$ the mass of particle and $a_{s}$ the scattering length.
For numerical computation, the above equation is made dimensionless by normalizing the
wave function and choosing
$6E_{R}$ as energy unit with $E_{R}=\hbar^2 k_L^2/2m$,  $ma^{2}/\pi^{2}\hbar$ as the
time unit,  and $\sqrt{3} a/2\pi$ as the length unit. The scaled nonlinearity and
potential strength are denoted as $c$ and $v$, respectively.

\begin{figure}
\includegraphics[bb=40 33 555 421,angle=0,scale=0.44]{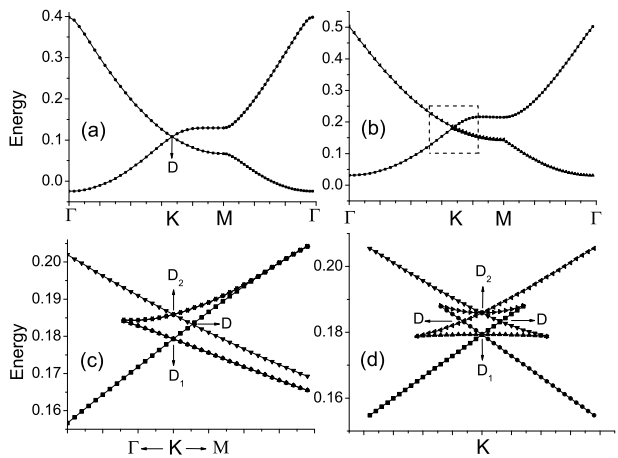}
\caption{ Bloch bands along the high symmetry points.
(a) The linear case. $c=0$, $v=0.1$. $D$ marks the Dirac point at point $\mathbf{K}$.
(b)The nonlinear case. $c=0.1$, $v=0.1$. (c) The enlarged rectangle part
in (b).  There appear two additional crossing points $D_1$ and $D_2$ while
the linear Dirac point $D$ is shifted away from $\mathbf{K}$.
(d) The band structure along the direction represented by the dashed line in Fig.\ref{hconfig}(b).}
 \label{fig1}
\end{figure}

We compute the Bloch wave solutions of the GP equation, which are of the form
$\psi_{\mathbf{k}}(\mathbf{r})=\sum_{m,n}c_{mn}e^{i(\mathbf{k+G}_{mn})\cdot \mathbf{r}}$
with  $\mathbf{G}_{mn}=m\mathbf{b}_{1}+n\mathbf{b}_{2}$,  and
the corresponding nonlinear Bloch bands. The bands along the high-symmetry point
are plotted in Fig.\ref{fig1}. Compared to the linear bands in
Fig.\ref{fig1}(a), we see that the nonlinear bands in (b) have a similar overall structure.
However, the part around  the Dirac point appears
to be modified by nonlinearity. When it is enlarged, we find in (c) that
the two linear bands have split into four bands. As a result, two more additional
crossing points $D_1$ and $D_2$ appear while the Dirac crossing $D$ is shifted
away from point $\mathbf{K}$. This feature is also clear in (d), where the Bloch band
along the direction 30$^\circ$ off the $\mathbf{K}$-$\mathbf{M}$ axis is plotted.
We have also plotted the nonlinear Bloch band
near point $\mathbf{M}$  in Fig.\ref{1d}(b) where we see a loop structure,
very similar to the BEC Bloch bands in one dimensional optical lattice\cite{Biaowu3,nlz,Machholm}.

\begin{figure}[!h]
\includegraphics[width=8cm]{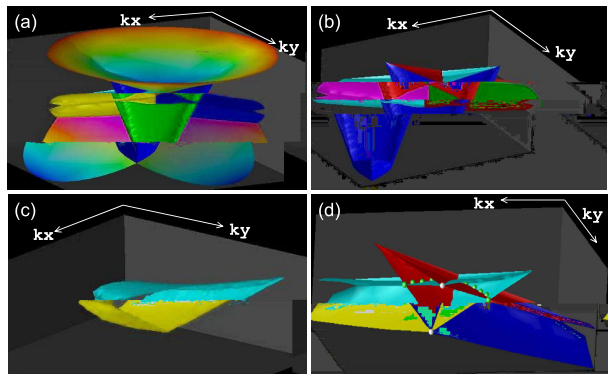}
\caption{(color) (a) The lowest nonlinear Bloch bands around point $\mathbf{K}$.
They consist of three intersecting ``tubes", which are sandwiched by two Dirac cones.
(b) Three intersecting ``tubes", which are aligned along the three $\mathbf{K}$-$\mathbf{M}$ axes;
(c) one of three ``tubes", whose cross-section area increases monotonically from $\mathbf{M}$
to $\mathbf{K}$;  (d) the cross-section of two intersecting ``tubes". The white dots are
$D_1$ and $D_2$ points in Fig.\ref{fig1}(c) while the green dots indicate
a closed curve where the $D$ point in Fig.\ref{fig1}(c) belongs.  $v=0.1, c=0.1$.}
 \label{hc3d}
\end{figure}

The full BEC Bloch bands near point $\mathbf{K}$ are plotted in Fig.\ref{hc3d}(a).
The complicated Bloch bands consist of three ``tubes", which intersect at
point  $\mathbf{K}$  (see Fig.\ref{hc3d}(b)) and are sandwiched by two Dirac cones.
One of the tubes is shown in Fig.\ref{hc3d}(c): it lies along the $\mathbf{M}$-$\mathbf{K}$
direction and it has a wedged cross-section with an area increasing monotonically
from $\mathbf{M}$ to $\mathbf{K}$ . Fig.\ref{hc3d}(d) shows how
two tubes intersect. The two white dots mark the top and bottom
tips of this intersection and correspond to $D_1$ and $D_2$ points in Fig.\ref{fig1}(c,d).
The green dots indicate part of a closed curve, which results from the intersection of the three tubes;
the shifted Dirac point $D$ in Fig.\ref{fig1}(c,d)
is one of the points on this closed curve. This shows that the Dirac point is turned into  a
closed curve by the interaction. The three tubes become smaller as the interaction
strength $c$ gets weaker. In particular, as $c$ decreases, the tube will disappear first
at point $\mathbf{M}$ and start shrinking toward point $\mathbf{K}$.
However, surprisingly, the tubes never disappear completely at $\mathbf{K}$ as long as $c$ is not zero.
This indicates that the tubed Bloch bands appear for arbitrary small interaction.
Note that the tips of the Dirac cones in Fig.\ref{hc3d}(a) have only triangular symmetry,
and  do not have the cylindrical symmetry as in the linear case.

\begin{figure}[!h]
\includegraphics[bb=20 38 566 395, angle=0,scale=0.40]{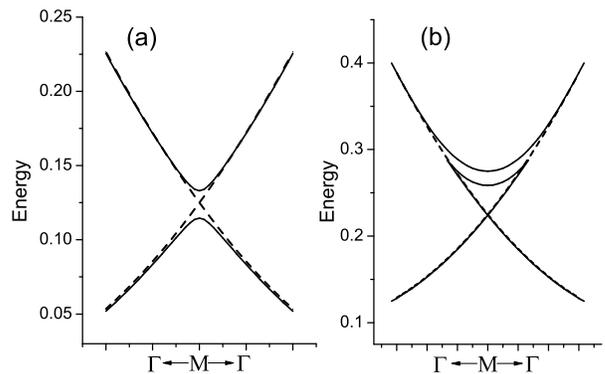}
\caption{The lowest Bloch bands for a honeycomb lattice at the
$\mathbf{\Gamma}$-$\mathbf{M}$-$\mathbf{\Gamma}$ branch.
(a) Free boson; (b) BEC. The solid curves are Bloch bands while the dashed curves
are the energy of the plane waves.}
 \label{1d}
\end{figure}
The tubed structure can be viewed as a fingerprint left in the Bloch bands by
the superfluidity of the BEC systems. Assume that we have
a mass flow of boson particles, which is represented by plane wave
$e^{i\mathbf{k}\cdot\mathbf{x}}$ with $\mathbf{k}$
at  the Brillouin zone (BZ) edge point $\mathbf{M}$. We now slowly turn on an
optical lattice of small lattice strength.  For free bosons, the flow is stopped by
the Brag scattering, the plane wave assumes the form of $\sin(\mathbf{k}\cdot\mathbf{x})$.
In the energy band, this is reflected by that the crossing of two plane wave energy
bands at  $\mathbf{M}$ is replaced by a gap as seen in Fig.\ref{1d}(a).
In the nonlinear case, the situation can be very different: when the interaction is strong
so that the superfluid critical velocity is larger than $|\mathbf{k}|$,  the small optical lattice, which
can be regarded as perturbation, should not stop the super-flow. This implies that the
wave function describing the flow should still resemble the plane wave
$e^{i\mathbf{k}\cdot \mathbf{x}}$, and at the same time, the crossing of plane wave energy bands should
remain unchanged. This is confirmed by our numerical calculation shown in Fig.\ref{1d}(b).
When this superfluidity argument is applied to other points along the BZ edge, we should have
a tubed structure seen in Fig.\ref{hc3d}. In other words, we can view the tubed structure
as the fingerprint left in the BEC Bloch bands by superfluidity.
For this hexagonal BEC system,
this fingerprint stays as long as $c$ is not zero.

We emphasize that the appearance of the tubed structure in the BEC Bloch bands for arbitrary small
$c$ is a unique feature for a honeycomb lattice: In one
dimensional lattice\cite{Biaowu3,nlz,Machholm,seaman,seaman2,Mueller,Larson} and
two dimensional square lattice\cite{chen2, chien}, the looped or tubed nonlinear structure in the Bloch bands
appears only when $c$ is bigger than a threshhold value.
This unique feature has a profound implication when the tight-binding limit is considered.
As is well known,  when the lattice is deep,
it is believed that the system should be well described by a tight-binding model\cite{smerzi}.
However, as shown below, the tight-binding model is not  an adequate approximation
for a BEC in a honeycomb lattice no matter how deep the lattice is.

Following the usual procedure\cite{haddad,smerzi}, we write the bosonic field as a sum
over the two sub-lattices $\psi=\sum_{\vec{a}}\psi_{\vec{a}}u(\vec{r}-\vec{a})+
\sum_{\vec{b}}\psi_{\vec{b}}u(\vec{r}-\vec{b})$,
where $u(\vec{r})$ is the Wannier function and $\vec{a}$ and $\vec{b}$ are lattice vectors
in the two sub-lattices, respectively. Then the tight-binding Hamiltonian for our BEC system is
\begin{equation} \label{dis}
H=-\sum_{<\vec{a},\vec{b}>}J_{\vec \delta}(\psi_{\vec{a}}^{*}\psi_{\vec{b}}+\text{h.c.}) +
\frac{U}{2}\Big[\sum_{\vec{a}}|\psi_{\vec{a}}|^{4}+\sum_{\vec{b}}|\psi_{\vec{b}}|^{4}\Big],
\end{equation}
where $J_{\vec \delta}$ is the hopping constant with ${\vec \delta}$ indicating three different neighbors
and $U$ is the on-site interaction proportional to $c$.  The ground state energy
of this Hamiltonian is $E=\pm|\Sigma_{1}|+U/2$, where
$\Sigma_{1}=-\sum_{\vec \delta}J_{\vec \delta}e^{i\mathbf{k}\cdot\vec \delta}$.
This shows that the interaction has only a trivial effect on the band structure, lifting
the Dirac bands by a constant $U/2$. This is very different from the GP equation result,
where arbitrary small interaction can destroy the Dirac bands.  This surprising difference
implies that the tight-binding model can not describe well the BEC system in a honeycomb
lattice no matter how deep the lattice is.

\begin{figure}[!h]
\includegraphics[bb=39 37 520 415, angle=0,scale=0.4]{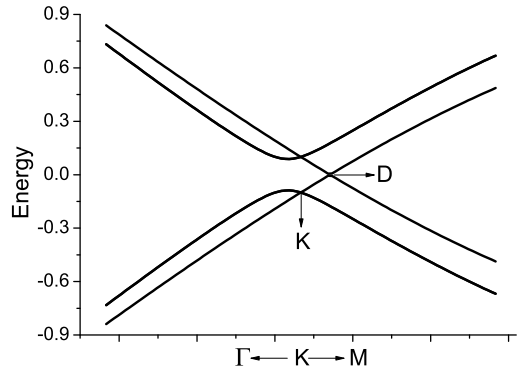}
\caption{The linear Bloch bands of the tight-binding model along the
$\mathbf{\Gamma}$-$\mathbf{M}$-$\mathbf{\Gamma}$ branch when
$J_{{\vec \delta}1}=J_{{\vec \delta}2}=J$ and $J_{{\vec \delta}3}=1.1J$.
The vertical axis is the energy in the unit of $J$.}
 \label{tba}
\end{figure}
After careful analysis, we find that the inadequacy of the tight-binding model may be
caused by the inappropriate choice of Wannier function. Let us consider the case
where $J_{{\vec \delta}1}=J_{{\vec \delta}2}=J$ and $J_{{\vec \delta}3}>J$\cite{jklee}.
The typical Bloch bands for this case are plotted in Fig. \ref{tba}, where one
immediately notices that the Dirac point is shifted and the bands at
point $\mathbf{K}$ are split, two important features that we also see in Fig.\ref{fig1}.
To have a different $J_{{\vec \delta}3}$, one needs a set of Wannier functions
which have no hexagonal rotational symmetry. This seems to suggest that
the choice of Wannier functions, which are used to obtain the tight-binding model, depend
on the state of the system.  Conventionally, the choice of Wannier functions
is independent of the state of the system. To confirm this, more computation
needs to be done and will be carried out in the future. One possible way of
doing the computation is to use the method proposed in Ref.\cite{WuShi}.


Interestingly,  the tubed structure shown in Fig.\ref{hc3d} has another important
physical implication, the breakdown of adiabaticity by nonlinearity.  In the linear case,
the state of the system can adiabatically follow the lower left band to the upper right band
by passing through the Dirac point.
In the nonlinear case, this adiabatic following is broken. This can be seen clearly
in Fig.\ref{hc3d}(d): the system can follow adiabatically passing point $D$ till the tip
of the band, where no more band to follow and the adiabaticity is broken.
This type of breakdown of adiabaticity is also implied in the loop structure in the
one dimensional optical lattice. Such an interesting effect has
not only been generalized to general nonlinear quantum systems\cite{jliu} but also
been observed experimentally with ultra-cold atoms\cite{yachen}.
However, the crucial difference of the hexagonal system is that  the breakdown of adiabaticity
occurs for arbitrary small interaction whereas it happens only when the nonlinearity is
bigger than a threshold value in the 1D system or other previously studied systems.

Note that this interesting phenomenon is not limited to the system of BEC in a honeycomb lattice.
It can be seen clearly when we approximate this BEC system at point $\mathbf{K}$
with a  three-mode model. The three-mode model is
\begin{widetext}
\begin{equation}
i\frac{\partial}{\partial t}\begin{pmatrix} \phi_1\\\phi_2\\\phi_3
\end{pmatrix}=\begin{pmatrix} -|\phi_{1}|^{2}c -\frac{\delta k_x}{4}-\frac{\sqrt{3}\delta k_y}{4}& v/2 & v/2 \\ v/2 & -|\phi_{2}|^{2}c+\frac{\delta k_x}{2}
 & v/2 \\ v/2 & v/2 & -|\phi_{3}|^{2}c -\frac{\delta k_x}{4}+\frac{\sqrt{3}\delta k_y}{4}\end{pmatrix}\begin{pmatrix} \phi_1\\\phi_2\\\phi_3\end{pmatrix}\,,
\end{equation}
\end{widetext}
where $\delta k_x$ and $\delta k_y$ denote how much the Bloch wave number $\mathbf{k}$
deviates away from $\mathbf{K}$.
This three-mode model can also describe a BEC in a triple-well potential,
where the three wells are arranged in a triangular geometry with the depth of
each well adjustable\cite{franzosi,jiang}.
It should also be realizable in experiment with waveguide systems\cite{Luo1,Szameit,anton}
and other nonlinear optical systems.
This shows that this breakdown of adiabaticity by arbitrary small interaction
is general and can happen in a wide range of systems.

Inspired by the experiment in \cite{yachen}, we here propose a scheme to realize the above mentioned triple well configuration. The procedure is as follows.  At first, a triangular lattice is formed by three lasers. The triangular
potential can be described by
$V_{\rm tri}=V_{0}\left [\cos(x+\frac{y}{\sqrt{3}})+\cos(-x+\frac{y}{\sqrt{3}})+\cos(\frac{2y}{\sqrt{3}})\right ]$
with $V_{0}<0$. The second step is to form the triple-well systems by adding a rectangular lattice  $V_{\rm rec}(\theta,\varphi)=V_{1}\left [\cos(x/2+\theta)+\cos(y/\sqrt{3}+\varphi)\right ]$ ($V_{1}>0$). As $\theta, \varphi$ changes, the second optical lattice can not only break the triangular lattice into a series of
independent triple-well systems but also change the depth of each well.  One should be
able to demonstrate the breakdown of adiabaticity by arbitrary small interaction with this triple-well system,
similar to the experiment done in Ref. \cite{yachen}.

In sum, we have computed the BEC Bloch bands in a honeycomb optical lattice. Our results show that
a tube-intersecting structure can emerge between the up and down Dirac cones for arbitrary small
interaction. This structure has two interesting physical implications: (1)
the tight-bind model can not describe adequately the BEC in a honeycomb lattice even when the lattice is very deep;
(2) the adiabaticity can be broken down by arbitrary small interaction in certain systems.  For the latter, we have
proposed an experimental scheme to observe it.

We thank Yongping Zhang for useful discussions. We also thank Zhaoxing Liang for the help on
numerical calculation. This work was  supported by the NSF of China (10825417).


\end{document}